\documentclass[aps,
twocolumn,
prd,dvipsnames,usenames,floatfix,superscriptaddress,nofootinbib,longbibliography]{revtex4-2}
\pdfoutput=1

\usepackage[utf8]{inputenc}
\usepackage{units}
\usepackage[english]{babel}
\usepackage{xspace}
\usepackage{amsmath} 
\usepackage{amssymb} 
\usepackage{graphicx}
\usepackage{hyperref}
\usepackage{color}
\usepackage{braket}

\usepackage{tikz}
\usetikzlibrary{positioning}
\usetikzlibrary {arrows.meta,bending,positioning}

\usepackage{xspace}
\newcommand*{\ie}{i.e., }
\newcommand*{\eg}{e.g., }
\newcommand*{\fig}{fig.\@\xspace}
\newcommand*{\eq}{eq.\@\xspace}
\newcommand*{\eqs}{eqs.\@\xspace}

\renewcommand{\Re}{\mathfrak{Re}\,}
\renewcommand{\Im}{\mathfrak{Im}\,}

\def\e{{\rm e}}

\usepackage{xifthen}
\newcommand*\diff{\mathrm{d}} 
\newcommand*\ldiff[2][]{ \ifthenelse{\isempty{#1}}{ \diff #2}{\diff^#1#2} \,} 
\let\limitint\int 
\renewcommand{\int}{\limitint \!} 

 \definecolor{backgroundcolor}{RGB}{0,26,128}

\begin{document}

\title{Non-Polynomial Interactions as a 
Path\\ Towards a Non-Renormalizable UV-Completion}

\begin{abstract}
We propose a new class of single-field scalar quantum field theories with non-polynomial interactions leading to a two-point Green's function that can be naturally continued beyond the naive cutoff scale. This provides a new prospect for self-completing theories in the UV-domain. In our approach, the formal power series for the scalar potential has a vanishing radius of convergence and is defined through Borel resummation. We discuss concrete examples, among others with a spectral function that vanishes at large momenta, potentially leading to an asymptotically free theory. Finally, we give an outlook on future research, with a view towards possible applications to inflation and gravity. 
\end{abstract}

	\author{Mikhail Shaposhnikov} 
	\email{mikhail.shaposhnikov@epfl.ch}
		\affiliation{Institute of Physics, 
		\'Ecole Polytechnique F\'ed\'erale de Lausanne, CH-1015 Lausanne, Switzerland}
	\author{Sebastian Zell}
	\email{sebastian.zell@uclouvain.be}
	\affiliation{Centre for Cosmology, Particle Physics and Phenomenology -- CP3,
		Universit\'e catholique de Louvain, B-1348 Louvain-la-Neuve, Belgium}

\maketitle

\newpage

\section{Possibilities for UV-completion} 
\label{intro}
The energy scales of Nature are distributed over a vast interval, ranging from neutrino mass differences and the density of dark energy on the order of $10^{-12}\,\text{GeV}$ to the Fermi scale $\sim 10^2\, \text{GeV}$ and finally the Planck mass $M_P\sim 10^{19}\, \text{GeV}$ set by gravity.  The energies that are reachable by the energy frontier experiments, most notably the Large Hadron Collider -- around $10^4\, \text{GeV}$ -- appear to be very low in comparison with $M_P$, meaning that direct observations of Planck scale physics are hardly possible. This represents one of the greatest challenges in fundamental physics.

Addressing this issue, one can wonder whether theories that have been tested at some low energy scale already contain information about their behavior at much larger energies. We can broadly distinguish three cases:
\begin{enumerate}
\item[(I)] First, there are \textit{renormalizable} or even finite theories that can be valid up to arbitrarily high energies. The most prominent representative in this category is the Standard Model (SM) of particle physics. With the experimentally determined parameter values, the SM remains consistent and weakly coupled up to the Landau poles in the Higgs and hyper-charge self-interactions, which only appear far above $M_P$.
\item[(II)] If a theory is nonrenormalizable, it features a cutoff scale $\Lambda$, above which it ceases to be predictive. This may signal the emergence of \textit{new heavy particles} with masses on the order of $\Lambda$. For example, the four-fermion theory of weak interactions becomes strongly coupled around $\Lambda \sim 10^2 \, \text{GeV}$.  For this case, a weakly coupled UV-completion is provided by the addition of new degrees of freedom contained in the Higgs boson and the intermediate vector bosons.
\item[(III)] A third category can exist between the two extreme scenarios outlined above. In this case, the theory becomes strongly coupled at some scale $\Lambda$. However, this does not signal the appearance of new physics but instead, a self-completion is achieved with the ingredients already present at low energies. This proposal of ``\textit{self-healing}'' was discussed in connection with Higgs inflation in \cite{Bezrukov:2010jz, Calmet:2013hia} and in a more general context in \cite{Aydemir:2012nz}, deriving its motivation from the theory of strong interactions (QCD), where the breakdown of chiral perturbation theory does not immediately correspond to the emergence of additional particles.
\end{enumerate}
	
The search for a UV-completion is particularly relevant in gravity. Certainly, General Relativity (GR) is not renormalizable, \ie does not belong to category (I). However, whether it corresponds to possibility (II) or (III) is a far-reaching open question. The former case is most prominently realised in supersymmetric and string theories, where many additional degrees of freedom emerge below $M_P$ (see \cite{Martin:1997ns,Hebecker:2020aqr} for reviews). The latter option of self-completion can \eg be implemented in scenarios of asymptotic safety \cite{Weinberg:1980, Reuter:1996cp, Berges:2000ew}, with applications to the Standard Model considered in \cite{Shaposhnikov:2009pv}, or via classicalisation \cite{Dvali:2010bf,Dvali:2010jz,Dvali:2011th}. 
	
In the present paper, we shall propose a new framework for constructing concrete self-completing theories in category (III). Our goal is to obtain non-renormalizable models that can be viewed as being ``close to renormalizable'', in the sense that IR-observables contain significant information about their UV-behavior. In addition to conceptual interest and possible implications for gravity, theories with non-polynomial potentials also appear in many theories of cosmological inflation, such as Starobinski inflation \cite{Starobinsky:1980te}, Higgs inflation \cite{Bezrukov:2007ep}, and $\alpha$-attractors models \cite{Kallosh:2013lkr,Kallosh:2013yoa}.
 For all of them, the possibilities of UV-completion still remain obscure, and our study may shed light on this problem. The cutoff $\Lambda$ in these theories is generically much smaller than the Planck scale (but much larger than the Fermi scale), allowing us to neglect gravity and to deal with a field theory in flat space.

\section{The general strategy}
\label{borel}
We shall use as an example a generic model of a massless scalar field $\varphi$ with  
\begin{equation} \label{Lagrangian}
\mathcal{L} = \frac{1}{2} \left(\partial_\mu \varphi\right)^2 - \lambda V(\varphi)\,,
\end{equation}
where $\lambda$ is a (small) coupling constant, the Minkowski metric $\eta_{\alpha\beta}$ has signature $(1,-1,-1,-1)$, and  $V(\varphi)$ is a generic non-polynomial function of $\varphi$, which can be {\em formally} expanded as power series in $\varphi$, 
\begin{equation} \label{potentialSeries}
V(\varphi) = \varphi^4 \sum_{m=0}^\infty V_m\left(\frac{\varphi}{\Lambda}\right)^m\,,
\end{equation}
where $V_m$ are dimensionless, $\Lambda$ is the cutoff of the theory and for simplicity we assumed the absence of superrenormalizable operators.

The contributions of interactions with $m>0$  give rise to a non-renormalizable theory. If the standard procedure of renormalization is used, one should introduce an infinite number of counter-terms and, therefore, the theory becomes unpredictive at energies exceeding the cutoff scale. 

Instead, to deal with this kind of potential one may consider perturbation theory for Green's functions (or S-matrix) with respect to the small coupling $\lambda$ and sum the entire perturbation series to all orders in $1/\Lambda$, as was proposed by Efremov and Fradkin in \cite{Efimov:1963yqa, Fradkin:1963vva} (for a review see \cite{Efimov:1969zz}).  It was found that for certain potentials the resummed perturbation theory may lead to a theory which is unitary, causal and UV-finite. Remarkably, the summation may even remove a number of the infinities, leading to a hope of having a self-consistent theory. This program was mainly developed in the 1960s having in mind the theory of strong interactions and gravity; see \cite{Efimov:1965aza, Lee:1969ni, Delbourgo:1969ppg, Salam:1970rp, Salam:1970jk, Lehmann:1971gq}  for some important works. 

The study of such theories decayed gradually after 1970 because renormalizable theories describing strong, weak and electromagnetic interactions were created and because technically the resummation is very difficult. These attempts were revisited recently by Magnin in his PhD thesis \cite{magnin} in connection with the UV-completion of Higgs inflation.  For potentials considered in these works, the cross-sections computed at second order in $\lambda$ increase unacceptably fast with energy (see also examples below), making the framework inconsistent in UV.\footnote{A further resummation, attempted in \cite{magnin}, may alleviate the problem, but cannot convincingly remove it.}

The crucial novelty of our approach is to consider potentials for which the power series \eqref{potentialSeries} has zero radius of convergence, but the potential $V(\varphi)$ makes sense via Borel resummation:
\begin{equation} \label{potentialBorel}
V(\varphi) = \varphi^4 \limitint_0^\infty \diff{t}\ \e^{-t} \sum_{m=0}^\infty \frac{V_m}{m!} \left(\frac{\varphi t}{\Lambda}\right)^m \;.
\end{equation}
As is well-known, $V(\varphi)$ coincides with the usual series \eqref{potentialSeries} if the latter exists, but we shall assume that this is not the case. Namely, we consider the scaling $V_m \sim \sqrt{m!}$ so that only \eq \eqref{potentialBorel} is well-defined. We shall call such potentials \textit{Borel-finite}. As described above,  we will not truncate the series in $m$, but perform a perturbative expansion in $\lambda$ by only taking into account the leading order in $\lambda^2$. In a sense, this is an approach inverse to that of resurgence (for a review see \cite{Dunne:2016nmc}), which uses the resummation of asymptotic series appearing in renormalizable field theories. In our case, the series for Green's functions are convergent (at least in some domain of momenta), but the interactions leading to them are represented by asymptotic series.

For the program to succeed, one should be able to compute {\em all} the elements of the S-matrix and demonstrate that they lead to a self-consistent field theory above the cutoff. Our goal in this work is much more modest. Namely,  we will be studying a simpler object, the propagator. In fact, from the behavior of the propagator at large energies, one can also extract information about the behavior of the vertex functions. In particular, if the two-point spectral function is positive (unitarity requirement) and vanishes at large momenta, the three-point vertex function also vanishes when the momentum transfer goes to infinity \cite{Barton1965, magnin} (a feature of asymptotically free field theories). It is conceivable to think that the same happens with higher vertex functions, though this has not been proven. An argument in favour of this possibility is the well-known result of axiomatic field theory that a theory is free if the exact propagator coincides with the free one \cite{Greenberg:1961mr, Bogolyubov:1975ps}.

The (momentum-space) two-point Green's function\footnote
{Here $\varphi$ represents full (Heisenberg) fields and $\mathbf{T}$ stands for time-ordering. With our conventions, the free propagator is 
\begin{equation*}
\braket{\mathbf{T} \varphi_{\text{free}}(x)\varphi_{\text{free}}(0)} = i \int \frac{\ldiff[4]{k}}{(2\pi)^4} \frac{\e^{-ikx}}{k^2+i\epsilon} \;,
\end{equation*}  
and its imaginary part can be computed using $\Im (k^2-i\epsilon)^{-1} = \pi \delta(k^2)$ (see \eg \cite{Bogoliubov1982}).
}
\begin{equation} 
\label{GreensFunctionDefinition}
G(k^2) =	i	\int \ldiff[4] x \e^{ikx}\,	\braket{\mathbf{T} \varphi(x)\varphi(0)} \;,
\end{equation}
can be equivalently expressed in terms of the spectral density (see \cite{Barton1965} for a review)
\begin{equation} \label{spectralDensityDefinition}
G(k^2) = \limitint_0^\infty \!\! \diff k^{'2} 					     
\frac{\rho(k^{'2})}{k^{'2}-k^2-i\epsilon}\,, \quad	\rho(k^{2}) = \frac{1}{\pi} \Im G(k^2) \,,
\end{equation}
where $\Im$ denotes the imaginary part. The cutoff scale splits
\begin{equation}
G(k^2) = \begin{cases}
G_{\text{IR}}(k^2) & k^2\leq \Lambda^2 \\
G_{\text{UV}}(k^2) & k^2 > \Lambda^2
\end{cases} \;,
\end{equation}
and analogously for $\rho(k^2)$.
	
At sufficiently small momenta $k^2<\Lambda^2$ (and at order $\lambda^2$ as throughout), we can compute $G_{\text{IR}}$ in terms of the potential \eqref{potentialBorel}. It turns out that this uniquely defines its imaginary part $\rho_{\text{IR}}$. In contrast, the real part depends on a priori unknown counter-terms from renormalization, expressible as functions $r_m$ to be defined later. Once this ambiguity contained in $r_m$ is fixed at low energies, this fully specifies a UV-completion $G_{\text{UV}}$ of the Green's function at momenta $k^2>\Lambda^2$. This is due to the fact that $G(k^2)$ must be analytic everywhere in the complex plane of $k^2$ except on the real demi-axis with $k^2>0$ (as is also evident from \eq \eqref{spectralDensityDefinition}). Clearly, $G_{\text{UV}}$ defines the imaginary part $\rho_{\text{UV}}$. In other words, the  Green's function cannot be directly computed from the potential for $k^2>\Lambda^2$ but it arises from analyticity, which leads to a resemblance with bootstrap methods \cite{Poland:2018epd,Arkani-Hamed:2018kmz, Kruczenski:2022lot}.

\begin{figure}[tbh]
\begin{tikzpicture}
		\node [draw,
		minimum width=1.3cm,
		]  (rhoUV) {$\rho_{UV}$};
		
		\node [draw,
		minimum width=1.3cm, 
		right=2.3cm of rhoUV
		] (GUV) {$G_{UV}$};
		
		\node [draw, 
		minimum width=1.3cm, 
		below = 1.5cm of rhoUV
		]  (rhoIR) {$\rho_{IR}$};
		
		\node [draw, 
		minimum width=1.3cm, 
		below = 1.5cm of GUV
		]  (GIR) {$G_{IR}$};
		
			\node [draw,
		minimum width=1.3cm, 
		left=2.3cm of rhoIR
		] (pot) {$V$};
		
		\draw[-{Stealth[width=3mm]}] (GUV.west) -- (rhoUV.east) 
		node[midway,above]{$\Im$};
		
		\draw[-{Stealth[width=3mm]},dashed,color=backgroundcolor] (rhoIR.east) -- (GIR.west)
		node[midway,above]{\textbf{\textcolor{backgroundcolor}{{Completion}}}}
			node[midway, below]{(\eg $r_m=0$)};
		
		\draw[-{Stealth[width=3mm]}] (pot.east) -- (rhoIR.west)
		node[midway, above]{perturbative}
			node[midway, below]{$O(\lambda^2)$};
		
		\draw[-{Stealth[width=3mm]}] (GIR.north) -- (GUV.south) 
		node[midway, left, align=right]{Analyticity};		
	\end{tikzpicture}
\caption{Schematic depiction of our approach. First, the Borel-finite potential $V$ defines the low-energy spectral density $\rho_{\text{IR}}$ perturbatively in the coupling $\lambda$. Then additional information is needed for also specifying the real part of the low-energy Greens function $G_{\text{IR}}$ (\eg from requiring the absence of additional contributions from renormalization, $r_m=0$). Once $G_{\text{IR}}$ is known, also $G_{\text{UV}}$ (and its imaginary part $\rho_{\text{UV}}$) are uniquely defined.}

\label{fig:scheme}
\end{figure}
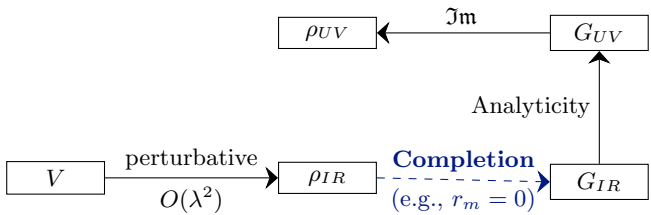

In comparison with the previous works \cite{Efimov:1963yqa, Fradkin:1963vva,magnin}, our approach is innovative in two aspects. The first one is the use of Borel-finite potentials. As we shall show, they may lead to $\rho_{\text{IR}}$ with a better UV-behavior, allowing for a UV-completion to second order in $\lambda$.  With certain assumptions specified below, a non-trivial imaginary part of $G_{\text{IR}}$ allows us to single out a specific (real part of) $G_{\text{IR}}$, which then uniquely defines the whole $G_{\text{UV}}$. This continuation of the Green's function from the IR to the UV represents our second innovation. Our approach for UV-completion from Borel-finite potentials is summarized in \fig \ref{fig:scheme}.  

In the remainder of the manuscript, we first compute in section \ref{sec:spectralFunction} the spectral function in terms of the potential. In section \ref{sec:prototypeModel}, we then present a concrete example of a Borel-finite potential and its UV-completion by analytic continuation. We give an outlook to future research in section \ref{sec:outlook} and conclude in section \ref{sec:conclusion}. Appendices \ref{app:SpectralDensity} and \ref{app:fourierTransform} contain details about the computations. In appendix \ref{app:toyModel}, we make a comparison to a renormalizable theory and appendix \ref{app:continuations} shows alternative parameter choices and UV-completions for our prototype model. 
		
\section{Computation of spectral function} 
\label{sec:spectralFunction}
We begin by deriving the low-energy Green's function $G_{\text{IR}}$ from a generic Borel-finite potential \eqref{potentialBorel}. Eq.\@\xspace \eqref{GreensFunctionDefinition} gives to second order in $\lambda$ (see details in appendix \ref{app:SpectralDensity})
\begin{equation} \label{GIRStart}
	\begin{split}
G_{\text{IR}}(k^{2}) = \frac{\lambda^2}{k^4} \sum_m |V_m|^2 \Lambda^{-2m} \, (m+4)\, (m+4)!\\ 
\cdot\int \ldiff[4] x \e^{ikx} \, i\  \mathcal{D}(x)^{m+3} \;,
\end{split}
\end{equation}
where $\mathcal{D}(x) =	\braket{\mathbf{T} \varphi_{\text{free}}(x)\varphi_{\text{free}}(0)}= -1/(4 \pi^2(x^2-i\epsilon))$ is the propagator of a massless scalar field in position space (see \cite{Bogoliubov1982}) and we left out the contribution of the free field, $\rho_{\text{free}}(k^2) = \delta(k^2)$. Eq.\@\xspace \eqref{GIRStart} equivalently holds for a potential with a usual power series expansion \eqref{potentialSeries}.

As shown in appendix \ref{app:fourierTransform}, we can explicitly calculate the Fourier transform in \eq \eqref{GIRStart}:
\begin{align}
	G_{\text{IR}}(k^{2}) & = \frac{\tilde{\lambda}^2}{\Lambda^2}  \sum_{m} f(m)\left(\frac{k^2}{\Lambda^2} \right)^m \nonumber\\
	&\left(r_m - \ln \left(\frac{-k^2 - i \epsilon}{\Lambda^2}\right)\right) \;. \label{GIR}
\end{align}
where we introduced $\tilde{\lambda} = \lambda/(4\pi)$ and the notation
\begin{equation} \label{modelPotentialCoefficients}
\sqrt{f(m)} =	 (-1)^{m} \frac{V_{m+1}\, (m+5)\sqrt{m+4}}{(4 \pi)^{m+2} \sqrt{(m+2)!}}  \;.
\end{equation}
Here $f(m)>0$ does not restrict the expansion coefficients $V_m$, up to a definite choice of alternating signs, which ensures convergence of the Borel resummation \eqref{potentialBorel}.
In \eq \eqref{GIR}, $r_m$ are real and a priori unknown numbers that arise from renormalization.
As already discussed before, this ambiguity due to $r_m$ reflects the fact that our theory is non-renormalizable.

We shall now discuss the resulting spectral density $\rho_{\text{IR}}=1/\pi\, \Im G_{\text{IR}}(k^{2})$. Within the radius of convergence of the series in $m$, \ie for $k^2 < \Lambda^2$, 
it is independent of the real contribution $r_m$ and hence determined uniquely.
First, we make a digression to the trivial case that $V_m=0$ for all $m\geq 1$. This corresponds to a renormalizable theory, \ie the limit $\Lambda\rightarrow \infty$ and $\rho \rightarrow \rho_{\text{IR}}$. Then it is evident from \eq \eqref{GIR} that $\rho(k^{2})$ vanishes as $k^2\rightarrow \infty$. We are interested in a self-completing scenario, in which the same behavior arises for non-zero and growing $V_m\sim \sqrt{m!}$. 

For a proof by contradiction, we momentarily assume that the radius of convergence of the series \eqref{GIR}, which defines $	G_{\text{IR}}(k^{2})$, is infinite. In this case, an imaginary part can only arise from the logarithm and the unknown contribution $r_m$ does not play a role. However, the summands in the remaining series (first line of \eq \eqref{GIR}) are manifestly positive and so the spectral function is unbounded at high energies, corresponding to catastrophic particle production, which contradicts our initial assumptions. Thus, we conclude that the radius of convergence cannot be infinite in \eq \eqref{GIR}. In principle, this still leaves open a possibility of an asymptotic series, meaning that not only the series expansion of the potential but also the Green's function \eqref{GIR} has a vanishing radius of convergence.
 We shall not consider this case and assume that the radius of convergence, and accordingly the cutoff scale $\Lambda$, are finite and non-zero.
In order to ensure that this property is fulfilled, we consider functions $f(m)$ that grow (or decay) at most exponentially for large $m$ so that \eq \eqref{modelPotentialCoefficients} yields the scaling $V_m\sim \sqrt{m!}$. 

\section{Prototype model}
\label{sec:prototypeModel}
\begin{figure}
	\centering
		\includegraphics[width=0.95\linewidth]{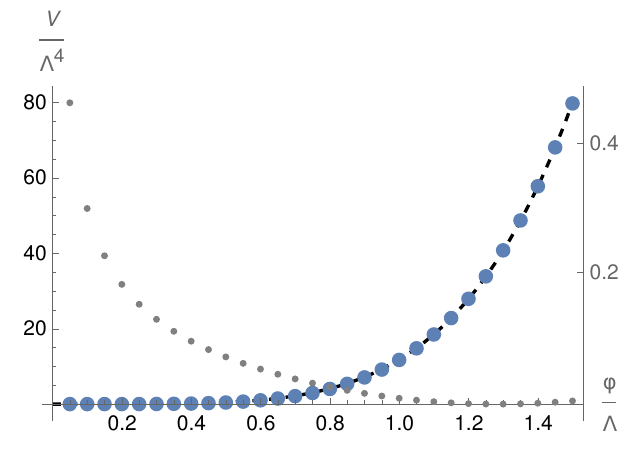}
	\caption{Potential $V(\varphi)$ as defined by \eq \eqref{fm} for $\alpha=9/10$ shown in blue dots. (For the numerical integration in \eq \eqref{potentialBorel}, we cutoff the sum in m at $23,000$ and the integration in $t$ at $90 M/\varphi$.) The dashed black line shows a fit by $f=a \varphi^b$ with $a \approx 12$ and $b\approx 4.7$. The relative error of the fit, $(f-V)/V$, is display as gray dots with axis on the right. We see that the fit is a good approximation for sufficiently large $\varphi$ but deteriorates around $\varphi=0$, where instead the usual power series \eqref{potentialSeries} can be used.}
	\label{fig:potential}
\end{figure}

Now we shall consider a concrete prototype model defined by
\begin{equation} \label{fm}
	f(m) = \begin{cases}
		\displaystyle- \frac{\Gamma(m+1-\alpha)}{\Gamma(-\alpha) (m+1)!} & m \geq 0 \\
		0 & m<0
	\end{cases} \;,
\end{equation}
where $\Gamma$ is the gamma function, $0<\alpha<1$, and we note that $f(m)$ only grows as a polynomial for large $m$. Plugging \eq \eqref{fm} into \eqs \eqref{modelPotentialCoefficients} and \eqref{potentialBorel}, we can compute the potential. The result is shown in \fig \ref{fig:potential}. 

Now we turn to the Green's function $G_{\text{IR}}(k^{2})$. The series in \eq \eqref{GIR} converges for $k^2<\Lambda^2$. (This holds true even if the two Borel resummations displayed in \eq \eqref{rhoStartBorel} are taken into account.)
Thus, $\Lambda$ is the naive cutoff scale, beyond which correlators are no longer defined. Within the radius of convergence, we get
\begin{align} 
\tilde{G}_{\text{IR}}(k^{2})
	& = \frac{\tilde{\lambda}^2}{k^2} \left(1-\left(1-\frac{k^2}{\Lambda^2}\right)^\alpha\right)\nonumber\\
	&\left(r_m - \ln \left(\frac{-k^2 - i \epsilon}{\Lambda^2}\right)\right)\label{GreensFunctionModel} \;,
\end{align}
Its imaginary part is uniquely defined:
\begin{equation} 
	\rho_{\text{IR}}(k^2)= \frac{\tilde{\lambda}^2}{k^2} \left(1-\left(1-\frac{k^2}{\Lambda^2}\right)^\alpha\right)\theta(k^2)\label{spectralFunctionModel} \;.
\end{equation}
Crucially it is non-zero for arbitrarily small $k^2$. This behavior is different from that in renormalizable theory where a heavy particle with mass $\sim \Lambda$ was integrated out; see appendix \ref{app:toyModel} for a concrete example.

\begin{figure}
 	\centering
 	\includegraphics[width=\linewidth]{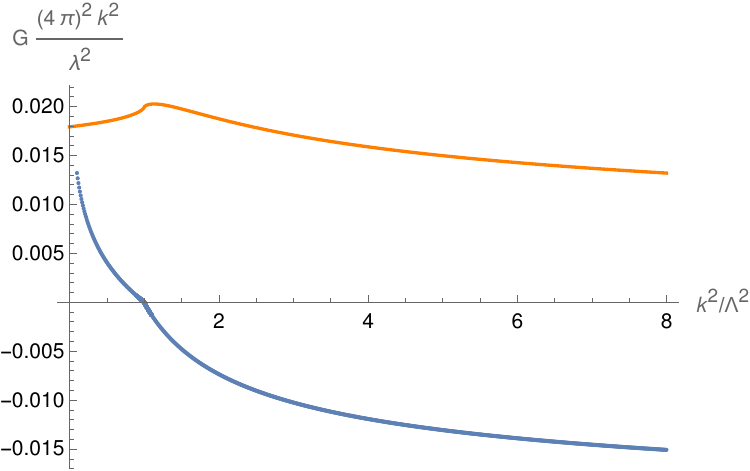}
 	\caption{Green's functions $G(k^2)$ for the prototype model \eqref{spectralFunctionModelContinued} for $\alpha=9/10$. The real part, $\Re G(k^2)$, and the imaginary part, $\Im G(k^2) = \pi \rho$, are shown in blue and orange, respectively.}
 	\label{fig:Green910}
 \end{figure}

As discussed, we shall define a UV-completion of the theory by specifying the real part of $G_{\text{IR}}(k^{2})$. As a first option, we consider $r_m=0$, corresponding to the possibility that no additional contributions arise due to renormalization. This uniquely defines 
\begin{align} 
	G^{(1)}_{\text{UV}}(k^{2})
	& = -\frac{\tilde{\lambda}^2}{k^2} \left(1-\left(1-\frac{k^2+i\epsilon}{\Lambda^2}\right)^\alpha\right) \ln \left(\frac{-k^2 - i \epsilon}{\Lambda^2}\right)\label{GreensFunctionModelContinued} \;,
\end{align}
and accordingly
\begin{align} 
	\rho^{(1)}_{\text{UV}}(k^{2})  &= \frac{\tilde{\lambda}^2}{k^2} \Bigg(1-\cos\left(\alpha \pi\right)\left(\frac{k^2}{\Lambda^2} -1\right)^\alpha    \nonumber\\
	& - \frac{1}{\pi}\sin\left(\alpha \pi\right)\left(\frac{k^2}{\Lambda^2} -1\right)^\alpha  \ln \left(\frac{k^2}{\Lambda^2} \right)    \Bigg)\; . \label{spectralFunctionModelContinued}
\end{align}
First, we observe that $\rho^{(1)}_{\text{UV}}(k^{2})$ will always become negative for sufficiently large $k^2$; we show the case $\alpha = 1/2$ in appendix \ref{app:continuations} (\fig \ref{fig:Green12}). For $\alpha$ sufficiently close to $1$, however, the scale at which $\rho^{(1)}_{\text{UV}}(k^{2})$ turns negative is at momenta $k^2\gtrsim \exp(1/(1-\alpha))\Lambda^2\gg \Lambda^2$.
 We display the result for $\alpha=9/10$ in \fig \ref{fig:Green910} (see also \fig \ref{fig:Spectral910} for larger values of $k^2$). Therefore, $	G^{(1)}_{\text{UV}}(k^{2})$ can still be viewed a partial UV-completion, with effects of new physics only becoming relevant at scales much larger than the naive cutoff $\Lambda$. 
It is important to note that the behavior of the spectral density \eqref{spectralFunctionModelContinued} is quite different from what one would expect when a threshold with the creation of a new particle opens up, see \fig \ref{fig:GreenToy}. Below $\Lambda$ our spectral density increases with energy, signalling the enhancement of particle production. Above $\Lambda$ it decreases, gradually switching off the particle production. We might also conjecture that the non-analyticity of $\rho^{(1)}_{\text{UV}}(k^{2})$ at  $k^2=\Lambda^2$ smears away when higher order corrections in $\lambda$ are accounted for.
 
  Finally, we remark that \eq \eqref{spectralFunctionModelContinued} provides a second approach for continuing the theory above the cutoff: One can simply take \eq \eqref{spectralFunctionModelContinued} as a definition of the spectral function without any reference to \eq \eqref{GreensFunctionModelContinued}.
In this spirit, we consider an alternative UV-completion with non-zero $r_m$, which is defined by
\begin{align}
\rho^{(2)}_{\text{UV}}(k^{2})   &= -\frac{\tilde{\lambda}^2}{\pi k^2} \Im \Bigg[ \left(1-\left(1-\frac{k^2-i\epsilon}{\Lambda^2} \right)^\alpha\right)\nonumber\\
&\ln \left(\frac{-k^2-i\epsilon}{\Lambda^2}\right)\Bigg] \; , \label{spectralFunctionAlternative}
\end{align}
and leads to replacing $\sin(\alpha \pi) \rightarrow - \sin(\alpha \pi)$ in \eq \eqref{spectralFunctionModelContinued}. Because of this modification, the spectral density now is positive, $\rho^{(2)}_{\text{UV}}(k^{2})>0$, for all values of $k^2$ and $\alpha$. Moreover, we have $\rho^{(2)}_{\text{UV}}(k^{2}) \rightarrow 0$ for $k^2\rightarrow \infty$, potentially leading to an asymptotically free field theory. 
 We show the resulting Green's function for $\alpha=1/2$ in appendix \ref{app:continuations} (\fig \ref{fig:GreenPositive}).

\section{Next steps}
\label{sec:outlook}
So far, we have only taken the very first steps in defining Borel-finite theories. In particular, the following directions of future research emerge from our results:
\begin{itemize}
	\item The next step will be to evaluate higher correlators beyond the two-point function. 
	\item So far, we have not performed an explicit renormalization, relating $r_m$ to counterterms.
	\item In particular, it remains to be determined if fixing $r_m$ suffices for uniquely specifying higher-point functions.
\item It would be interesting to consider a case in which $\sum_m f(m) k^{2m}$ grows exponentially but this growth is canceled by a suitably chosen $r_m$.
\item Finally, our goal is to construct UV-complete inflationary scenarios and models with gravitational interactions.
\end{itemize}

\section{Conclusion}
\label{sec:conclusion}

Understanding non-renormalizable theories above their cutoff scale is of utmost importance, in particular for finding a UV-completion of GR. In this paper, we have proposed a new approach for importing some of the advantageous properties of renormalizable theories, in which the low-energy domain determines the high-energy regime, to the non-renormalizable case. Our key ingredient are potentials that have a vanishing radius of convergence but are instead defined through Borel resummation. Although at first sight, this makes the theory more complicated, we have shown that resulting Green's functions possess a natural continuation to high energies and therefore open up a new perspective for self-completion. This has a twofold implication -- both for some of the leading inflationary models and for gravity itself.

\begin{acknowledgments}
 We thank Giorgos Karananas for insightful feedback on the manuscript. The work of M.S.~was supported by the Generalitat Valenciana grant PROMETEO/2021/083.  S.Z.~acknowledges support of the Fonds de la Recherche Scientifique -- FNRS.
\end{acknowledgments}

\appendix

\section{Computation of spectral density}
\label{app:SpectralDensity}
We shall compute \eq \eqref{GreensFunctionDefinition} to second order in $\lambda$:
\begin{align} 
G_{\text{IR}}(k^{2}) &= 	\int \ldiff[4] x \e^{ikx} \,	i (-\frac{1}{2}\lambda^2) \int \diff^4 z_1 \int \diff^4 z_2 \nonumber\\
	& \limitint_0^\infty \diff t_1 \limitint_0^\infty \diff t_2 \, \text{e}^{-t_1 - t_2}		 \sum_{n=0}^\infty \sum_{m=0}^\infty \frac{V_n V_m t_1^n t_2^m}{n! m!\Lambda^{n+m}}  \nonumber\\
	& \braket{\mathbf{T} \varphi(x)\varphi(0)\varphi^{n+4}(z_1) \varphi^{m+4}(z_2)} \;.
\end{align}
Evaluating the time-ordered product with Wick's theorem, we see that only summands with $m=n$ contribute. Representing the two propagators that involve the external positions $x$ and $0$ in momentum space, we carry out two of the spatial integrals to arrive at
\begin{align} 
G_{\text{IR}}(k^{2}) & = \frac{\lambda^2}{k^4} \limitint_0^\infty \diff t_1 \limitint_0^\infty \diff t_2 \, \text{e}^{-t_1 - t_2}  \sum_m \frac{(t_1 t_2)^m |V_m|^2}{m!^2 \Lambda^{2m}}\nonumber \\
& \, (m+4)! (m+4)\, \int \ldiff[4] x \e^{ikx} \, i\, \mathcal{D}^{m+3}(x) \;, \label{rhoStartBorel}
\end{align}
where $\mathcal{D}(x)$ is the position space propagator. For sufficiently small $k^2$, we can pull the Borel integrals over $t_1$ and $t_2$ inside the sum over $m$, which yields \eq \eqref{GIRStart}.

\section{Computation of Fourier transform} 
\label{app:fourierTransform}
In order to compute the Fourier transform in \eq \eqref{GIRStart}, we employ dimensional regularization by analytically continuing to $D$ dimensions \cite{tHooft:1972tcz}. Moreover, we go to Euclidean space, $x_0\rightarrow -i x_4$ (see \eg \cite{Weinberg:1995mt}), where the propagator becomes $\mathcal{D}_E(x) = 1/(4 \pi^2 x_E^{D-2})$. Thus, the quantity of interest is mapped as follows: 
\begin{equation}
	\int \ldiff[4]{x} \e^{ikx} \, i \, \left(\frac{1}{x^2-i \epsilon}\right)^{m+3} \rightarrow  \int \ldiff[D]{x_E} \e^{-ipx} \left(\frac{1}{x_E^{D-2}}\right)^{m+3} \;,
\end{equation}
where $p$ denotes the Euclidean momentum and we shall drop the subscript $E$. In full analogy to the computation  in \cite{Chetyrkin:1980}, we can calculate
\begin{align}
	&\int \ldiff[D]{x} \e^{-ipx} \left(\frac{1}{x^{D-2}}\right)^{m+3}\nonumber \\
	= &  \frac{\pi^{D/2} \Gamma\left(\frac{D}{2} - (m+3)(\frac{D}{2}-1)\right)}{\Gamma\left(\frac{D}{2}\right)\Gamma\left(2-\frac{D}{2} + (m+3)(\frac{D}{2}-1)\right)} \left(\frac{p^2}{4}\right)^{(m+2)(D/2-1) -1} \;.  \label{fourierFormula}
\end{align}
For any given value of $m$, the above integral converges if $D$ is sufficiently close to $2$. Now we continue back analytically to $D=4-2\epsilon$, which yields
\begin{align}
	&\int \ldiff[D]{x} \e^{-ipx} \left(\frac{1}{x^{D-2}}\right)^{m+3}\nonumber\\
	& = \frac{\pi^{2-\epsilon}\, \Gamma\left(-(m+1)+\epsilon(m+2)\right)}{\Gamma\left(2-\epsilon\right)\Gamma\left(m+3 -\epsilon(m+2)\right)} \left(\frac{p^2}{4}\right)^{m+1-\epsilon(m+2)}\nonumber\\
	& = \frac{(-1)^{m+5}\pi^{2}}{((m+2)!)^2} \left(\frac{p^2}{4}\right)^{m+1} \times\\ &\qquad \left(\epsilon^{-1}  + g_m \right) \left(1 -\epsilon(m+2)\ln \frac{p^2}{4}\right) + O(\epsilon)\;,
\end{align}
where we used in the last step that $m\geq0$ and $\Gamma\left(-(m+1)+\epsilon(m+2)\right)=\epsilon^{-1} (-1)^{m+1}/(m+2)! + O(\epsilon^0)$. Moreover, we defined
\begin{equation}
	g_m =  1-\gamma -\ln \pi +(m+2)(\Psi(m+2) + \Psi(m+3)) \;.
\end{equation}
Here $\Psi$ is the polygamma function and we note that $g_m>0$ for $m\geq 0$. 
After continuing back to Lorentzian signature, $-p^2 \rightarrow k^2 + i \epsilon$, we arrive at \eq \eqref{GIR}.

\section{Renormalizable toy model}
\label{app:toyModel}

As a simple example of the usual power series expansion, we consider in addition to $\varphi$ another scalar field $\chi$ of mass $M$:
\begin{equation} \label{toyModel}
	\mathcal{L} = \frac{1}{2} \left(\partial_\mu \varphi\right)^2 + \frac{1}{2} \left(\partial_\mu \chi\right)^2 - \frac{M^2}{2} \chi^2 - g \varphi \chi^2  \;.
\end{equation}		
We shall compute correlators of the massless $\varphi$-field to leading order in $g^2$. Only the 2-point function is non-trivial and reads
\begin{equation} \label{fullPropagatorToy}
G(k^2) = i 	\int \ldiff[4] x \e^{ikx}\,	\braket{\mathbf{T} \varphi(x)\varphi(0)} = \frac{2 g^2}{\left(k^2\right)^2} F(k^2) \;,
\end{equation}
where the momentum space representation of the loop is
\begin{equation}
	F(k^2) = \frac{-i}{(2\pi)^4} \int \ldiff[4]k^{'2} \frac{1}{k^{'2}-M^2 + i \epsilon}\, \frac{1}{(k' + k)^2-M^2 + i \epsilon} \;.
\end{equation}
In $\overline{\text{MS}}$-renormalization, we get \cite{Brown:1992db}
\begin{align} \label{Fren}
	F(k^2) &= \frac{-1}{(4 \pi)^2} \Bigg(\ln \left(\frac{M^2}{4\pi\mu^2}\right)\nonumber\\
	&+\limitint_0^1 \diff \alpha \frac{\alpha(1-2\alpha)}{\frac{M^2}{k^2 + i \epsilon} - \alpha(1-\alpha)}\Bigg) \;,
\end{align}
where $\mu$ is a mass scale.

We can compute explicitly\footnote
{We use that for real $c$
\begin{equation}
	\int \diff \alpha \frac{1}{\alpha^2 -\alpha +c - i \epsilon} = \begin{cases}
		\frac{2}{\sqrt{4c-1}} \arctan\left(\frac{2\alpha - 1}{\sqrt{4c-1}}\right) & c >1\\
		\frac{1}{\sqrt{1-4c}} \ln\left(\frac{1-\frac{2 \alpha -1}{\sqrt{1-4c}} + i \epsilon}{1+\frac{2 \alpha -1}{\sqrt{1-4c}}}\right) & c<1
	\end{cases} \;.
\end{equation}
}
\begin{equation} \label{FrenIntegrated}
	F(k^2) = \frac{-1}{(4 \pi)^2} \left(\ln \left(\frac{M^2}{4\pi\mu^2}\right) -2 + \tilde{F}(k^2) \right)\;,
\end{equation}
where
\begin{equation} \label{FTilde}
 \tilde{F}(k^2) = \begin{cases}
 	\displaystyle	2\sqrt{\frac{4 M^2}{k^2} - 1}\ \arctan\sqrt{\frac{1}{\frac{4 M^2}{k^2} - 1}} & k^2 < 4M^2 \\
 \displaystyle	- \sqrt{1-\frac{4M^2}{k^2}}\  \ln \frac{\sqrt{1-\frac{4M^2}{k^2}}-1 + i \epsilon}{\sqrt{1-\frac{4M^2}{k^2}}+1}	 &  k^2 > 4M^2
 \end{cases} \;.
\end{equation}

\begin{figure}
		\centering
			\includegraphics[width=\linewidth]{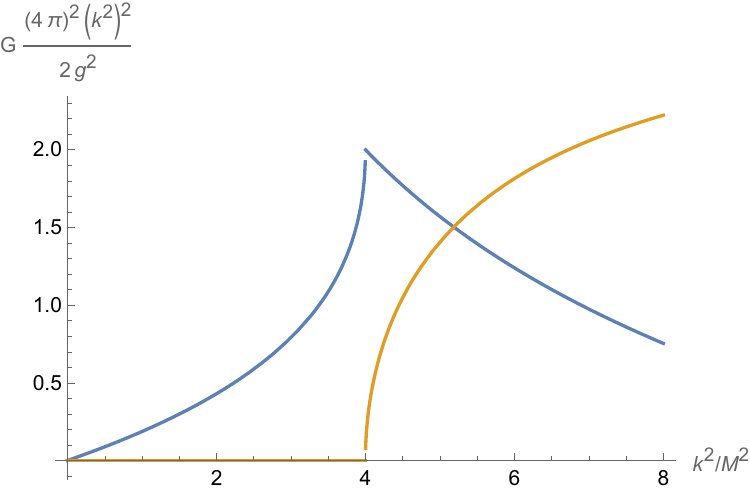}
	\caption{Green's functions in toy model defined by \eqs \eqref{fullPropagatorToy} and \eqref{FTilde}. The real part, $\Re G^{(2)}(k^2)$, and the imaginary part, $\Im G^{(2)}(k^2) = \pi \rho^{(2)}(k^2)$, are shown in blue and orange, respectively. }
	\label{fig:GreenToy}
\end{figure}

From here we can calculate the spectral density. Applying \eq \eqref{spectralDensityDefinition}, we obtain
\begin{eqnarray} \label{spectralDensityToyAppendix}
	\rho(k^{'2}) = \frac{2 g^2}{(4\pi)^2 (k^{'2})^2} \theta(k^{'2} - 4 M^2)\sqrt{1 - \frac{4 M^2}{k^{'2}}} \;.
\end{eqnarray}
As important difference to the prototype model defined by \eq \eqref{fm}, $\rho(k^{'2})$ vanishes for $k^{'2} < 4 M^2$. We can also investigate if the real part of \eq \eqref{FTilde} is continuous. Close to $k^2=4M^2$, we can approximate to leading order
\begin{equation} 
\Re	\tilde{F}(k^2) \approx \begin{cases}
			\pi\sqrt{\frac{4 M^2}{k^2} - 1} & k^2 < 4M^2 \\
		2 \left(1-\frac{4M^2}{k^2}\right) &  k^2 > 4M^2
	\end{cases} \;.
\end{equation}
Thus, the derivative is
\begin{equation} \label{discontinuousDerivative}
\frac{\partial\, \Re	\tilde{F}(k^2)}{\partial k^2} \approx \begin{cases}
		-\frac{\pi}{8M^2\sqrt{\frac{4 M^2}{k^2} - 1}} \rightarrow - \infty & k^2 < 4M^2 \\
	\frac{1}{2}	 &  k^2 > 4M^2
	\end{cases} \;.
\end{equation}
Thus, the 2-point function is continuous but not differentiable. We display the resulting Green's function in \fig \ref{fig:GreenToy}.

In a second approach, we shall compute the integral in \eq \eqref{Fren} in Euclidean signature by continuing $k^2+i\epsilon \rightarrow -k_\text{E}^2$. This yields a fully convergent integrations resulting in \eq \eqref{FrenIntegrated} with
\begin{equation} \label{FTildeEuclidean}
	 \tilde{F}(k_\text{E}^2) = \sqrt{1+\frac{4M^2}{k_\text{E}^2}}\  \ln \frac{\sqrt{1+\frac{4M^2}{k_\text{E}^2}}+1}{\sqrt{1+\frac{4M^2}{k_\text{E}^2}}-1} \;.
\end{equation}
This result shows explicitly that the Euclidean Green's function is analytic.
Now we continue back to Lorentzian momentum, $k_\text{E}^2 \rightarrow -k^2 - i \epsilon$. For $k^2 < 4M^2$, we have $\sqrt{1-\frac{4M^2}{k^2+i \epsilon}} = \sqrt{1-\frac{4M^2}{k^2} + i \epsilon\, \text{sgn}(k^2)} = \text{sgn}(k^2) i \sqrt{\frac{4M^2}{k^2} -1}$, which yields
\begin{equation}
	\tilde{F}(k^2<4M^2)  = i \sqrt{\frac{4M^2}{k^2} -1} \ \ln \frac{i\sqrt{\frac{4 M^2}{k^2} - 1}+1}{i\sqrt{\frac{4 M^2}{k^2} - 1}-1} \;.
\end{equation}
Since $\arctan 1/x = i/2 \ln\left((ix+1)/(ix-1)\right)$ for real $x$, this coincides with the first line of \eq \eqref{FTilde}. Now for $k^2>4M^2$, we use $\sqrt{1-\frac{4M^2}{k^2+i \epsilon}} = \sqrt{1-\frac{4M^2}{k^2}} + i \epsilon$ so that we obtain
\begin{equation}
	\tilde{F}(k^2>4M^2)  = \sqrt{1-\frac{4M^2}{k^2}}\  \ln \frac{\sqrt{1-\frac{4M^2}{k^2}}+1}{\sqrt{1-\frac{4M^2}{k^2}}-1 + i \epsilon}  \;,
\end{equation}
in accordance with the second line of \eq \eqref{FrenIntegrated}. So the Euclidean computation yields the same result as evaluating the 2-point function in Lorentzian signature.

\clearpage

\section{Plots of UV-completions}
\label{app:continuations}
\begin{figure}[b]
	\centering
	\includegraphics[width=\linewidth]{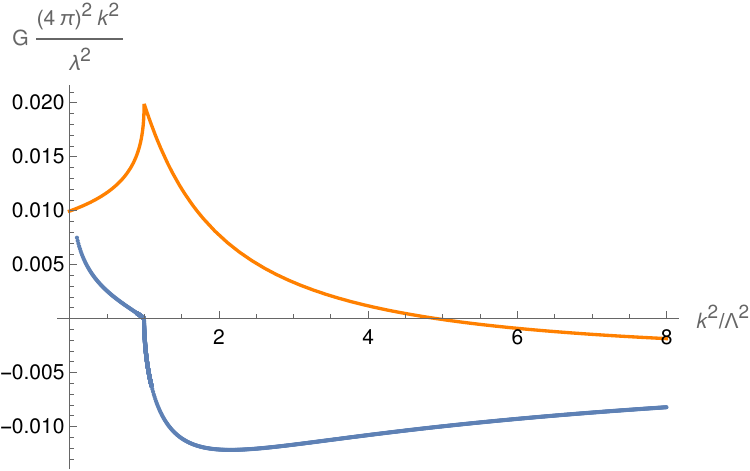}
	\caption{Green's functions $G^{(1)}(k^2)$ for the prototype model \eqref{spectralFunctionModelContinued} for $\alpha=1/2$. The real part, $\Re G^{(1)}(k^2)$, and the imaginary part, $\Im G^{(1)}(k^2) = \pi \rho^{(1)}(k^2)$, are shown in blue and orange, respectively. We see that the spectral density becomes negative, at which point the theory ceases to be valid and needs to be supplemented by contributions from new physics.}
	\label{fig:Green12}
\end{figure}

\begin{figure}[b]
	\centering
	\includegraphics[width=\linewidth]{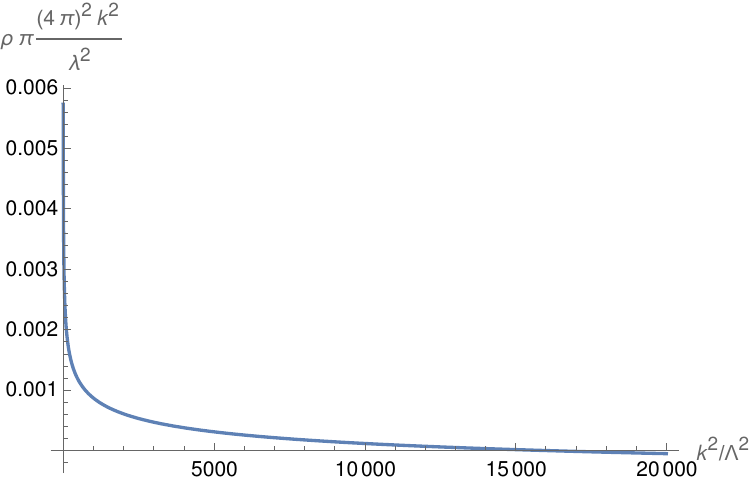}
	\caption{Spectral density $\rho^{(1)}(k^2)$ for the prototype model \eqref{spectralFunctionModelContinued} for $\alpha=9/10$. We observe that $\rho^{(1)}(k^2)$ only becomes negative for very large $k^2\gtrsim 10^4 \Lambda^2$.}
	\label{fig:Spectral910}
\end{figure}

\begin{figure}[b]
		\centering
		\includegraphics[width=\linewidth]{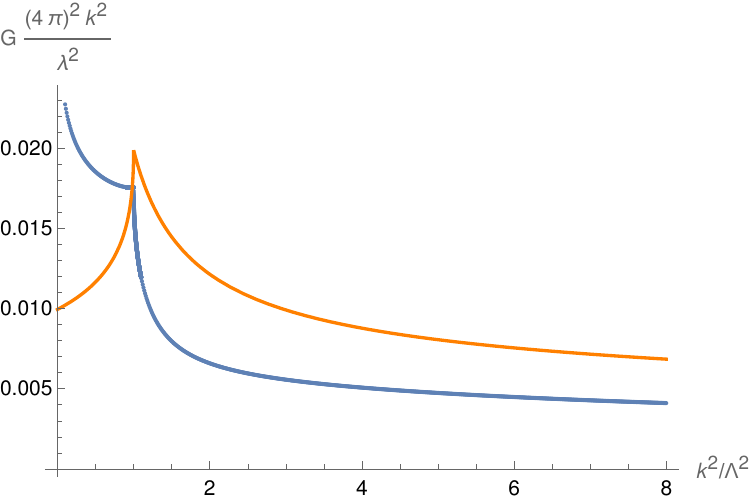}
	\caption{Green's functions $G^{(2)}(k^2)$ for the alternative model defined by \eq \eqref{spectralFunctionAlternative} for $\alpha=1/2$. The real part, $\Re G^{(2)}(k^2)$, and the imaginary part, $\Im G^{(2)}(k^2) = \pi \rho^{(2)}(k^2)$, are shown in blue and orange, respectively. Now $\rho^{(2)}(k^2)$ is positive for all $k^2$.}
	\label{fig:GreenPositive}
\end{figure}

In this appendix, we show some supplementary plots for section \ref{sec:prototypeModel}. Figs.\@\xspace \ref{fig:Green12} and \ref{fig:Spectral910} refer to the UV-completion defined by \eq \eqref{GreensFunctionModelContinued} while \fig \ref{fig:GreenPositive} show a result for the alternative continuation in \eq \eqref{spectralFunctionAlternative}.

\clearpage

\clearpage
\bibliography{Refs}

\end{document}